\documentclass[letterpaper, 10 pt, conference]{ieeeconf}  

\usepackage{amsmath}
\usepackage{tikz}
\usepackage{amssymb}
\usepackage{array}
\usepackage{graphicx}

\renewcommand{\thesection}{\textbf{\large \roman{section}}}

\IEEEoverridecommandlockouts                              
\title{\LARGE \fontfamily{cmss} \selectfont 
\textbf{Register Aggregation for Hardware Decompilation}
}

\author{\large
Varun Rao$^{1\dag}$, Zachary Sisco$^2$\\
\small
$^1$Mission San Jose High School\\
$^2$Department of Computer Science, University of California, Santa Barbara\\
$^\dag$corresponding author: varrao23@gmail.com
}

\begin{document}

\maketitle
\thispagestyle{empty}
\pagestyle{empty}

\section*{\raggedright \textbf{ABSTRACT}}
\noindent Hardware decompilation reverses logic synthesis, converting a gate-level digital electronic design, or netlist, back into hardware description language (HDL) code. Existing techniques decompile data-oriented features but do not address memory elements, which pose difficulty due to their deconstruction into data flip-flops in netlists and the cycles they form. Recovering multi-bit registers and memory blocks expands the applications of hardware decompilation, notably towards retargeting technologies (ie FPGAs to ASICs) and decompiling processor memories. We devise a method for register aggregation, to identify relationships between the flip-flops in a netlist, categorize them into registers and memory blocks, and output HDL code instantiating these memory elements. We group flip-flops by common enable pins and derive register bit-orders using functional dependencies, scaling up similarly to two-dimensional memory blocks. We evaluate our technique over a dataset of netlists, comparing the quantity and widths of the recovered registers and memory blocks with the netlist source code. The technique successfully recovers memory elements in all of the tested circuits, even aggregating beyond the source code expectation. In 10/13 circuits, all source code memory elements are accounted for, and we are able to compact up to 2048 disjoint bits into a single memory block. \\

\noindent \textbf{Keywords:} hardware decompilation, hardware description languages, sequential logic, electronic design automation

\section{\raggedright \textbf{INTRODUCTION}}
Hardware description languages (HDLs) are often employed for designing large-scale digital electronics, given their large inventory of abstractions. They take advantage of logic synthesis tools to produce netlists, which are graphical circuit representations consisting of individual logical gates that prove useful during manufacturing. Since synthesis distills HDL code down into gates, the semantic meaning of the code gets lost in the synthesized netlist, making these netlists quite difficult to analyze and interpret. Synthesis is also nondeterministic, involving numerous optimizations, incompleteness, and occasional errors [\ref{b1}–\ref{b3}]. Therefore, hardware designers must run extensive simulations to confirm that their HDL code and the netlist synthesized from it have identical behavior [\ref{b4}].

The problem of hardware decompilation has been proposed to reduce this simulation time and compress netlists for easier use. Hardware decompilation produces HDL code deterministically from a netlist by recognizing various abstractions and collapsing them into code. Netlists are much larger artifacts than their code counterparts [\ref{b5}], so decompiling a netlist into HDL code allows it to be simulated much faster. Beyond improving simulation speed, hardware decompilation has a number of other applications, such as netlist compaction, transpilation between HDLs, and propagating netlist edits back up to code. Existing work in hardware decompilation deals with features like loops [\ref{b6}] and modules [\ref{b7}], which are more prevalent in combinational logic. However, these methods can only decompile sequential circuits in certain contexts, lacking a robust way to deal with larger memory elements.

In this paper, we address hardware decompilation with respect to the defining feature of sequential logic: registers and memory blocks. The gates forming a netlist are bitwise operators, so any registers present in the netlist must be split into single-bit data flip-flops (DFFs) during synthesis, losing their connection to one another. 

We aim to perform register aggregation, to recover the multi-bit registers and memory blocks in a netlist which were originally instantiated in the equivalent HDL code.  This contributes significantly to processor decompilation given that registers, and especially memory blocks, are key elements of almost all processors. Additionally, register aggregation allows for technology retargeting, such as from field-programmable gate arrays (FPGAs) to application-specific integrated circuits (ASICs), once memory blocks can be manipulated as cohesive pieces of the netlist. There is existing work in forming multi-bit registers in a netlist, but this is applied to reverse engineering instead of decompilation and focuses primarily on data flow, neglecting bit-order and not outputting HDL code [\ref{b8}]. It also does not support memory block recovery. Other reverse engineering work does produce HDL code but does not target memory elements [\ref{b9}-\ref{b10}].

We propose a two-step approach to aggregate the DFFs in a netlist into registers: first dividing them into groups and then ordering each group to form a multi-bit register. We follow a similar approach in the second dimension for memory blocks. We group DFFs by enable, and derive their relative order according to functional relationships between DFFs. This is inspired by a characterization of sequential logic using linearly inductive boolean functions [\ref{b11}]. The key contributions of this paper are:
\begin{itemize}
\item We describe and implement a technique to group and order DFFs into registers (Section 2A).
\item We describe and implement a technique to scale our work up to two-dimensional memory blocks by aggregating multi-bit registers (Section 2B).
\item We evaluate our technique on a set of benchmark netlists with known memory elements by comparing the sizes and quantities found (Section 3). 
\end{itemize}

\section{\raggedright \textbf{METHODS}}

\subsection{Aggregating Registers \label{2A}}
We begin by partitioning the set of  netlist DFFs into register groups, which we accomplish through grouping DFFs by their enable inputs. This is motivated by the idea that DFFs belonging to the same multi-bit register should all be activated by the same enable signal. Grouping by enable limits our algorithm to considering only DFFs with an enable pin, which is reasonable given that hardware decompilation is an inherently heuristic process. 

Once the partition has been made, we search for an ordering of each register group. We do this by finding the register transfer arc of each DFF in the group, which is the set of nodes in the netlist that are ancestors of that particular DFF. We find the transfer arc of a DFF by performing an upward depth-first search from it on the netlist, searching specifically for other DFFs in the register group to find functional dependencies. A dependency relationship expresses that the input into one DFF is predicated upon the output from another. 

Key to the depth-first search is dismantling the cycles formed by the DFFs, which we do by separating each DFF into two nodes: one for its current value and one for the next value to be stored. We organize the dependencies into a directed graph, where the nodes are the DFFs in the register group, and each directed edge represents that one DFF is dependent upon another. Note how DFF $y$ being dependent on DFF $x$ (denoted $x < y$) is directly transferable to $x$ coming before $y$ in the register’s ordering. This is because if there is no dependency between two bits in a register, their pairwise bit-order does not actually affect the circuit’s functionality; a dependency between two DFFs shows that the circuit relies on their pairwise order in the register. This means that our graph of $<$ dependencies is essentially a graph of pairwise orders. 

In the graph, a valid ordering consists of a sequence of the DFFs that obeys as many $<$ dependencies as possible (ideally all). This is a classic scenario where we can employ topological sort. Since the $<$’s form directed edges in the graph, a topological sort would naturally build an ordering that resolves all of the $<$ operations if such an ordering exists, or pick one which resolves as many $<$’s as possible. Oftentimes, there are not enough dependencies to leave the topological sort with only one valid ordering, leading it to pick a different ordering from the originally intended one. However, this is because the intended ordering was chosen in a partially arbitrary manner, so the modified portion of the computed ordering does not actually matter to the circuit’s output. Thus, we have an algorithm to partition the registers in a circuit and order each grouping, addressing our twofold goal for multi-bit registers. 

\begin{figure} 
\centering 
\begin{tikzpicture}[scale=0.9]
\def \x{0.22}

\node[thick,draw,align=center, minimum width = 25pt] (r0) at (0,2) {$r_0$};
\draw[thick] (0 - 0.3 * \x,2 - \x) -- (0,2 - 0.6 * \x);
\draw[thick] (0 + 0.3 * \x,2 - \x) -- (0,2 - 0.6 * \x);

\node[thick,draw,align=center, minimum width = 25pt] (r0n) at (0,-2) {$r_0.n$};
\draw[thick] (0 - 0.3 * \x,-2 - \x) -- (0,-2 - 0.6 * \x);
\draw[thick] (0 + 0.3 * \x,-2 - \x) -- (0,-2 - 0.6 * \x);

\node[thick,draw,align=center, minimum width = 25pt] (r1) at (3,2) {$r_1$};
\draw[thick] (3 - 0.3 * \x,2 - \x) -- (3,2 - 0.6 * \x);
\draw[thick] (3 + 0.3 * \x,2 - \x) -- (3,2 - 0.6 * \x);

\node[thick,draw,align=center, minimum width = 25pt] (r1n) at (3,-2) {$r_1.n$};
\draw[thick] (3 - 0.3 * \x,-2 - \x) -- (3,-2 - 0.6 * \x);
\draw[thick] (3 + 0.3 * \x,-2 - \x) -- (3,-2 - 0.6 * \x);

\node[thick,draw,align=center, minimum width = 25pt] (r2) at (6,2) {$r_2$};
\draw[thick] (6 - 0.3 * \x,2 - \x) -- (6,2 - 0.6 * \x);
\draw[thick] (6 + 0.3 * \x,2 - \x) -- (6,2 - 0.6 * \x);

\node[thick,draw,align=center, minimum width = 25pt] (r2n) at (6,-2) {$r_2.n$};
\draw[thick] (6 - 0.3 * \x,-2 - \x) -- (6,-2 - 0.6 * \x);
\draw[thick] (6 + 0.3 * \x,-2 - \x) -- (6,-2 - 0.6 * \x);

\node[thick,circle,draw,align=center] (not) at (0,-0.5) {$not$};
\node[thick,circle,draw,align=center] (xor1) at (3,-0.5) {$xor$};
\node[thick,circle,draw,align=center] (and) at (4.5, 0.5) {$and$};
\node[thick,circle,draw,align=center] (xor2) at (6, -0.5) {$xor$};

\node[align=center] (en) at (1.5, 3.3) {$en$};
\draw[thick] (1.5, 3) -- (2, 3.5) -- (1, 3.5) -- (1.5, 3);

\node[thick,circle,fill,align=center,inner sep=0pt,minimum size=4pt] (p1) at (0, 1.4) {};
\node[thick,circle,fill,align=center,inner sep=0pt,minimum size=4pt] (p2) at (2.6, 0.85) {};
\node[thick,circle,fill,align=center,inner sep=0pt,minimum size=4pt] (p3) at (3, 1.5) {};
\node[thick,circle,fill,align=center,inner sep=0pt,minimum size=4pt] (p4) at (1.5, 2.8) {};
\node[thick,circle,fill,align=center,inner sep=0pt,minimum size=4pt] (p5) at (1.5, -2) {};

\draw[thick] (1.5, 3) -- (p4);
\draw[thick] (p5) -- (p4);
\draw[thick] (p4) -- (7.2, 2.8);
\draw[thick] (7.2, -2) -- (7.2, 2.8);
\draw[thick] (r0) -- (p1);
\draw[thick] (p2) -- (p1);
\draw[thick] (r1) -- (p3);
\draw[thick,-{stealth}] (p1) -- (not);
\draw[thick,-{stealth}] (p2) -- (xor1);
\draw[thick,-{stealth}] (p2) -- (and);
\draw[thick,-{stealth}] (p3) -- (xor1);
\draw[thick,-{stealth}] (p3) -- (and);
\draw[thick,-{stealth}] (not) -- (r0n);
\draw[thick,-{stealth}] (and) -- (xor2);
\draw[thick,-{stealth}] (xor1) -- (r1n);
\draw[thick,-{stealth}] (xor2) -- (r2n);
\draw[thick,-{stealth}] (r2) -- (xor2);
\draw[thick,-{stealth}] (p5) -- (r0n);
\draw[thick,-{stealth}] (p5) -- (r1n);
\draw[thick,-{stealth}] (7.2, -2) -- (r2n);

\draw[line width=0.17mm,color=red] (0,-1.6) -- (1.5,2.3) -- (-1.5,2.3) -- (0,-1.6);
\draw[line width=0.17mm,color=green] (3,-1.6) -- (4.6,2.7) -- (-1.5,2.7) -- (3,-1.6);
\draw[line width=0.17mm,color=blue] (6,-1.6) -- (7.5,2.5) -- (-1.9,2.5) -- (6,-1.6);

\end{tikzpicture}
\caption{The gate-level netlist for a 3-bit counter with an enable pin, with the corresponding register transfer arcs for each data flip-flop drawn in different colors.}
\end{figure}
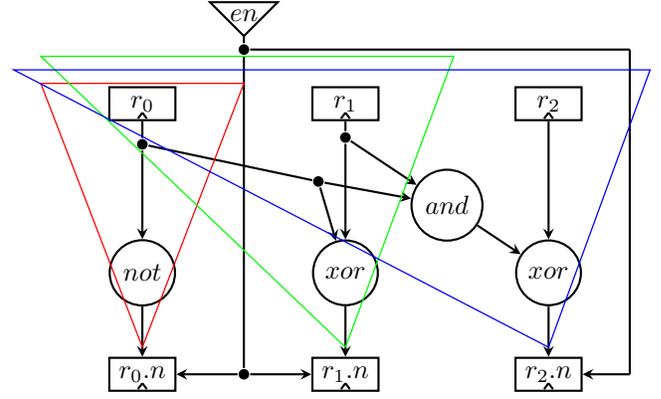

To illustrate our method, we provide an example visualizing the ordering technique on a 3-bit counter netlist in Figure 1, assuming the partition has already identified the 3 DFFs as a register group. The register transfer arcs for the 3-bit counter can be formalized into equations, with respect to a high enable bit.
If the superscript $n$ denotes the $n$th clock cycle, we have
\begin{equation}\label{eq1}
\begin{array}{rl}
&r_0^n = \neg r_0^{n-1},\\[5pt]
&r_1^n = r_0^{n-1} \oplus r_1^{n-1},\\[5pt]
&r_2^n = ( r_0^{n-1} \land r_1^{n-1} ) \oplus r_2^{n-1}
\end{array}
\end{equation}

Equation \ref{eq1} can be generalized to an inductive formula [8] for all counter DFFs:

\begin{equation}\label{eq3}
\begin{array}{rl}
&r_i^0 = 0,\\[5pt]
&r_i^n = (r_0^{n-1} \land r_1^{n-1} \land \dots \land r_{i-1}^{n-1} ) \oplus r_i^{n-1}
\end{array}
\end{equation}

Extracting functional relationships between DFFs from equation \ref{eq3} yields:

\begin{equation}\label{eq2}
r^n_0(r^{n-1}_0), r^n_1(r^{n-1}_0, r^{n-1}_1), r^n_2(r^{n-1}_0, r^{n-1}_1, r^{n-1}_2)
\end{equation}

\begin{figure}
\centering
\begin{tikzpicture}[scale=0.6]
\node[thick,circle,draw=none,align=center] at (-1,2) {a.};
\node[thick,circle,draw,align=center] (r0) at (0,2) {$r_0$};
\node[thick,circle,draw,align=center] (r1) at (2,2) {$r_1$};
\node[thick,circle,draw,align=center] (r2) at (4,2) {$r_2$};

\draw[thick,-{stealth}] (r1) -- (r0);
\draw[thick,-{stealth}] (r2) -- (r1);
\draw[thick,-{stealth}] (r2) to[out=140,in=40] (r0);
\draw[thick,-{stealth}] (r0) to[out=-60,in=-120, looseness=3] (r0);
\draw[thick,-{stealth}] (r1) to[out=-60,in=-120, looseness=3] (r1);
\draw[thick,-{stealth}] (r2) to[out=-60,in=-120, looseness=3] (r2);

\node[thick,circle,draw=none,align=center] at (-1,0) {b.};
\node[thick,circle,draw,align=center] (r0) at (0,0) {$r_0$};
\node[thick,circle,draw,align=center] (r1) at (2,0) {$r_1$};
\node[thick,circle,draw,align=center] (r2) at (4,0) {$r_2$};

\draw[thick,-{stealth}] (r1) -- (r0);
\draw[thick,-{stealth}] (r2) -- (r1);
\draw[thick,-{stealth}] (r0) to[out=-60,in=-120, looseness=3] (r0);
\draw[thick,-{stealth}] (r1) to[out=-60,in=-120, looseness=3] (r1);
\draw[thick,-{stealth}] (r2) to[out=-60,in=-120, looseness=3] (r2);

\node[thick,circle,draw=none,align=center] at (-1,-2) {c.};
\node[thick,circle,draw,align=center] (r0) at (0,-2) {$r_0$};
\node[thick,circle,draw,align=center] (r1) at (2,-2) {$r_1$};
\node[thick,circle,draw,align=center] (r2) at (4,-2) {$r_2$};

\draw[thick,-{stealth}] (r0) to[out=-60,in=-120, looseness=3] (r0);
\draw[thick,-{stealth}] (r1) to[out=-60,in=-120, looseness=3] (r1);
\draw[thick,-{stealth}] (r2) to[out=-60,in=-120, looseness=3] (r2);
\end{tikzpicture}
\caption{The dependency graph for the 3-bit counter in Figure 1 is topological sorted into an ordering in (a). The dependency graphs for a 3-bit bitwise operation (b) and a 3-bit shifter (c) are also shown for comparison.
}
\end{figure}
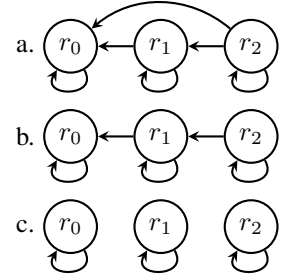

The relationships stated in \ref{eq2} are then used to form a directed graph of dependencies between the DFFs, upon which topological sort is performed, creating the final ordering, as shown in Figure 2a. The dependency graph for a counter shows each DFF being dependent on all of the DFFs coming after it in the register, but for other small building blocks we find different dependency relationships. For example, the dependencies in a shifter (Figure 2b) create relationships only between consecutive bits instead. For a circuit where the register value is updated using a bitwise operation with an input wire, each of the DFFs is completely independent, so all possible orderings are valid and will preserve the integrity of the circuit  (Figure 2c). 

\subsection{Aggregating Memory Blocks \label{2B}}
After aggregating multi-bit registers, we wish to form memory blocks by grouping suitable registers of the same width into a matrix of DFFs. Under the assumption that the memory blocks support enabled writes, we can begin with grouping by enable again, this time by searching for a wire which is a common input into the respective enables of each multi-bit register in the memory group. Unlike the register scenario, there is no need to find a vertical ordering between the registers in the memory group because they will be deleted and replaced with the memory block instantiation. However, memory blocks are more abstract than multi-bit registers, with defined ports beyond just an organized collection of DFFs. To form a complete memory block, we also need to identify and bundle the wires corresponding to each port by searching around the registers in the memory group. We consider memory blocks with the following four ports: one read port, one write port, one address, and an enable. 

We have already described how to find the enable. To find the address, we move upward on the netlist from the registers in search of the logic that selects which register to write to based on the bits of the address. This logic is recognizable in the netlist because it consists solely of ands and nots in terms of the single-bit wires corresponding to each address bit: these are the wires that must be bundled into the address. Since each register has a unique address, some unique subset of the address bits is negated and conjuncted with the rest of the address bits to form the address. We leverage the fact that every register in the memory group will have these ands and nots in terms of the same address bits, allowing us to find the address reliably. 

Since we assume that there is only one write port, all the registers will have their DFF data inputs updated by the same set of wires when the enable is high, which is the write port. Finally, for the read port, we follow a multiplexer tree stemming from the registers until it stops at a series of $n$ wires, where $n$ is the width of each register. The multiplexers are synthesized due to choosing which register is read from given the address, with the select pin of each multiplexer being an address bit. We infer the depth of the multiplexer tree from the address width to find the read port wires. Bundling all the ports completes memory block aggregation: we can now instantiate memory blocks using the widths of their registers and addresses, as well as their bundled input and output ports.

We implement our full aggregation technique entirely in Python, inputting netlists as Berkeley Logic Interchange Format (BLIF) [\ref{b12}] files and then parsing them to a more usable format in the PyRTL HDL [\ref{b13}]. Our output modifies the PyRTL working block, which stores all the netlist wires and gates, for easy translation to code. 

\section{\raggedright \textbf{EVALUATION} \label{3}}

\begin{figure}[t] \label{data}
\centering
\resizebox{0.485\textwidth}{!}{
\begin{tabular}{ |>{\hspace{-3pt}}m{74 pt}|m{30 pt}|>{\hspace{-3pt}}m{28 pt}|>{\hspace{-3pt}}m{19 pt}|m{76 pt}|>{\hspace{-3pt}}m{18 pt}|m{43 pt}| } 
 \hline
 Name \newline & \centering HDL \newline & \centering $n_\text{gates}$ \newline & \centering $n_\text{reg}$ \newline & \centering Register metadata \newline (size : quantity) & \centering $n_\text{mem}$ \newline & Memory \newline dimensions  \\ 
 \hline
 \hline
 alu & \centering PyRTL & \centering 23432 & \centering 0 & \centering & \centering 1 & $16 \times 32$ \\
 \hline
 fifo & \centering PyRTL & \centering 2227 & \centering 3 & 3 : 3 & \centering 1 & $8 \times 8$ \\
 \hline
 piso & \centering PyRTL & \centering 855 & \centering 6 & 1 : 1, 2 : 1, 4 : 4 & \centering 1 & $8 \times 8$ \\
\hline
bsg\_assembler & \centering Verilog & \centering 55564 & \centering 20 & 1 : 10, 16 : 10 & \centering 0 & \\
\hline 
bsg\_cache & \centering Verilog & \centering 542038 & \centering 4842 & 1 : 2759, 2 : 2,\newline 3 : 4, 4 : 2,\newline 8 : 2056, 16 : 2, \newline 28 : 2, 32 : 10,\newline 33 : 1, 40 : 1,\newline 64 : 1, 65 : 2 & \centering 0 & \\
\hline 
bsg\_fifo & \centering Verilog & \centering 98 & \centering 1 & 16 : 1 & \centering 0 & \\
\hline
bsg\_idiv & \centering Verilog & \centering 16753 & \centering 8 & 1 : 3, 32 : 2, \newline 33 : 3 & \centering 0 & \\
\hline
bsg\_lfsr & \centering Verilog & \centering 62 & \centering 1 & 1 : 1 & \centering 0 & \\
\hline 
bsg\_multiply\_array & \centering Verilog & \centering 4567 & \centering 15 & 1 : 3, 5 : 1, 9 : 1, \newline 13 : 1, 16 : 9 & \centering 0 & \\
\hline
bsg\_multiply & \centering Verilog & \centering 11140 & \centering 12 & 5 : 1, 6 : 1, 7 : 1, \newline 8 : 1, 10 : 1,\newline 11 : 1, 12 : 1,\newline 14 : 1, 16 : 4 & \centering 0 & \\
\hline
bsg\_strobe & \centering Verilog & \centering 660 & \centering 0 & \centering & \centering 0 &\\
\hline
nerv & \centering Verilog & \centering 90836 & \centering 6 & 1 : 2, 5 : 2, 32 : 1 & \centering 1 & $32 \times 32$ \\
\hline
opdb\_pico & \centering Verilog & \centering 111066 & \centering 102 & 1 : 74, 2 : 4, 3 : 1, \newline 4 : 5, 5 : 3, 7 : 1,\newline 8 : 1, 12 : 1, \newline 25 : 1, 31 : 1, \newline 32 : 9, 64 : 1 & \centering 2 & $4 \times 32$, \newline $32 \times 32$ \\
\hline
\end{tabular}
}
\caption{Table displaying each benchmark, the HDL used, gate count, register count, register metadata, memory dimensions, and memory counts.}

\end{figure}

We test our method on a set of 13 benchmark circuits designed in Verilog (mainly from the Basejump STL [\ref{b14}]) and PyRTL [\ref{b10}] with known registers and memory blocks, described in Figure 3. The benchmarks consist mainly of circuit building blocks, like arithmetic operations, caches, and register buffers, but they also include two CPUs: nerv (a RISC V processor) and opdb\_pico from OPDB [\ref{b15}].

We compare the source code registers and memory blocks to the technique’s outputted predictions for each benchmark (see Figure 4 for an overview across benchmarks).  

\subsection{Categorization of the Main Study }

We find that the results can be categorized into groups. For the alu, fifo (Figure 5a), bsg\_fifo, and bsg\_lfsr benchmarks, our technique performs perfect aggregation, or behaves exactly as expected. It recovers all the enabled registers and memory blocks originally present in the source code with identical dimensions. Notably, we use bsg\_strobe as a negative example, where there are only DFFs without enables; the technique does not produce any false positives and outputs zero registers. 

\begin{figure*} \label{overview}
\centering
\begin{tikzpicture}[scale=1.0]
\node at (5,5) {\includegraphics[scale=0.4]{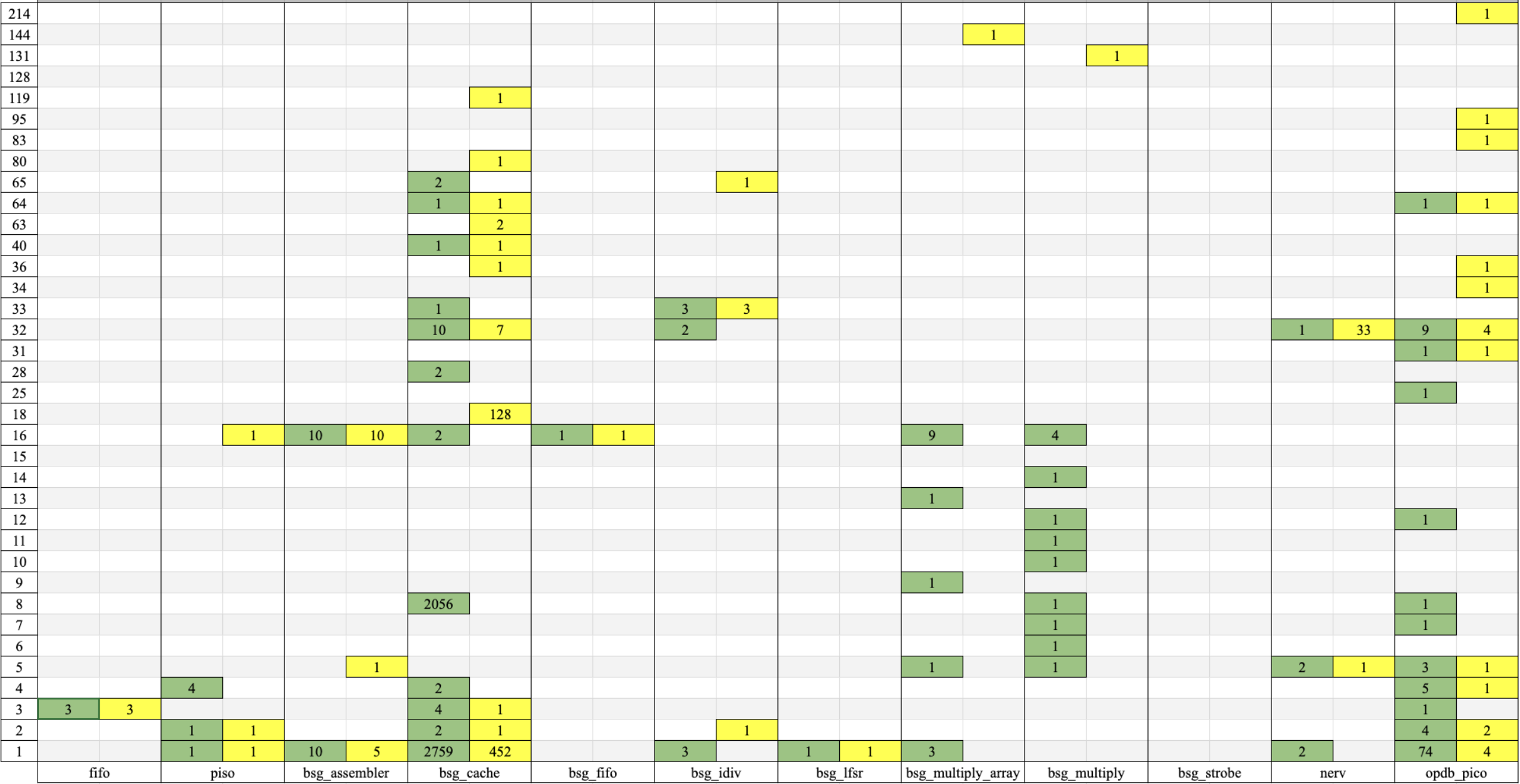}};
\node at (5,0.25) {\footnotesize Benchmark};
\node at (-3.8,5)[rotate=90] {\footnotesize Register Size};
\end{tikzpicture}
\caption{An overview of the register sizes and quantities for each benchmark. The source values are in green (same data from Figure 3), and the predictions outputted by the technique are in yellow.}
\end{figure*}

\begin{figure*} \label{overview}
\centering
\begin{tikzpicture}[scale=0.87]
\node at (5.9,0) {\includegraphics[scale=0.275]{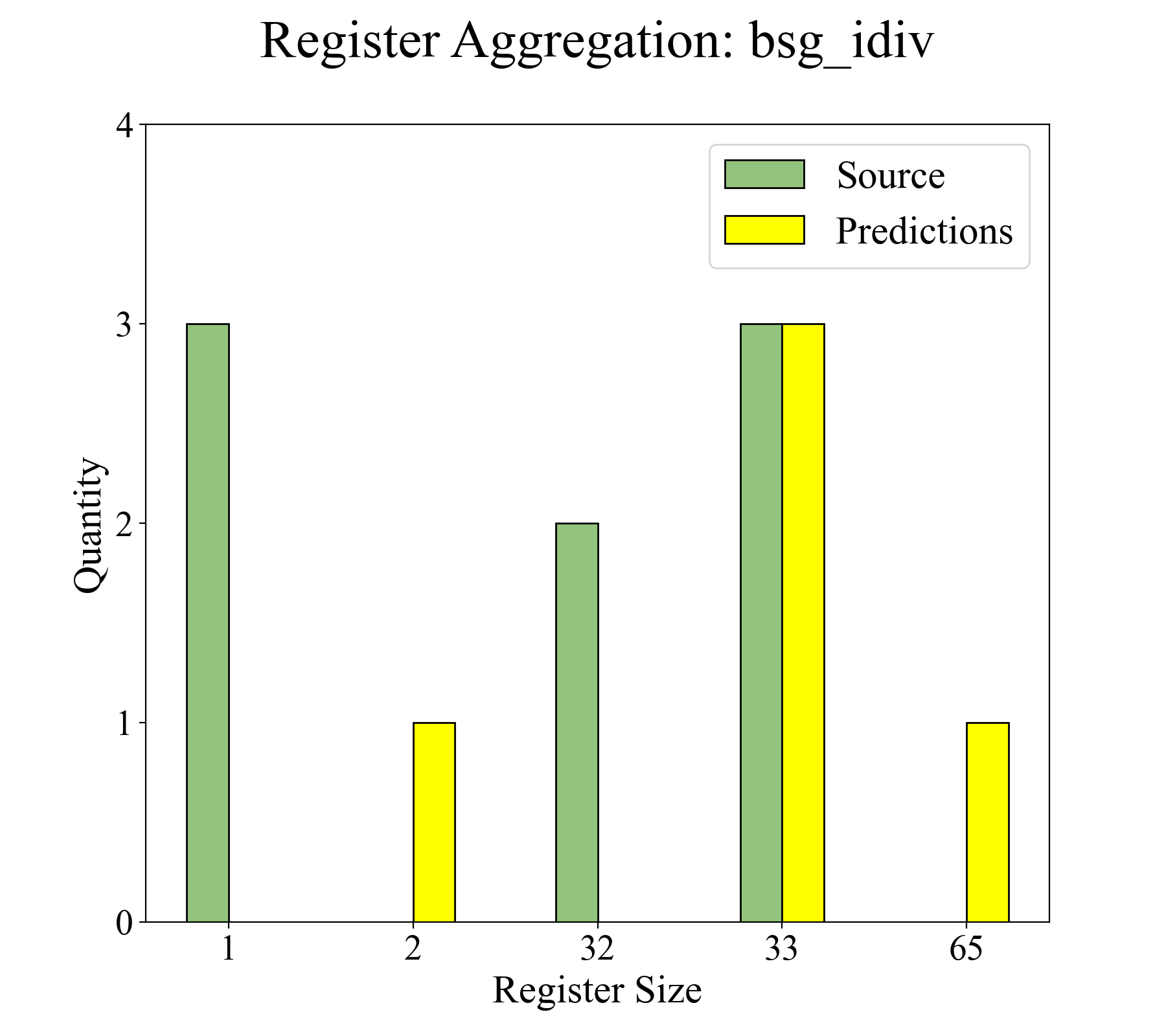}};
\node at (0,0) {\includegraphics[scale=0.275]{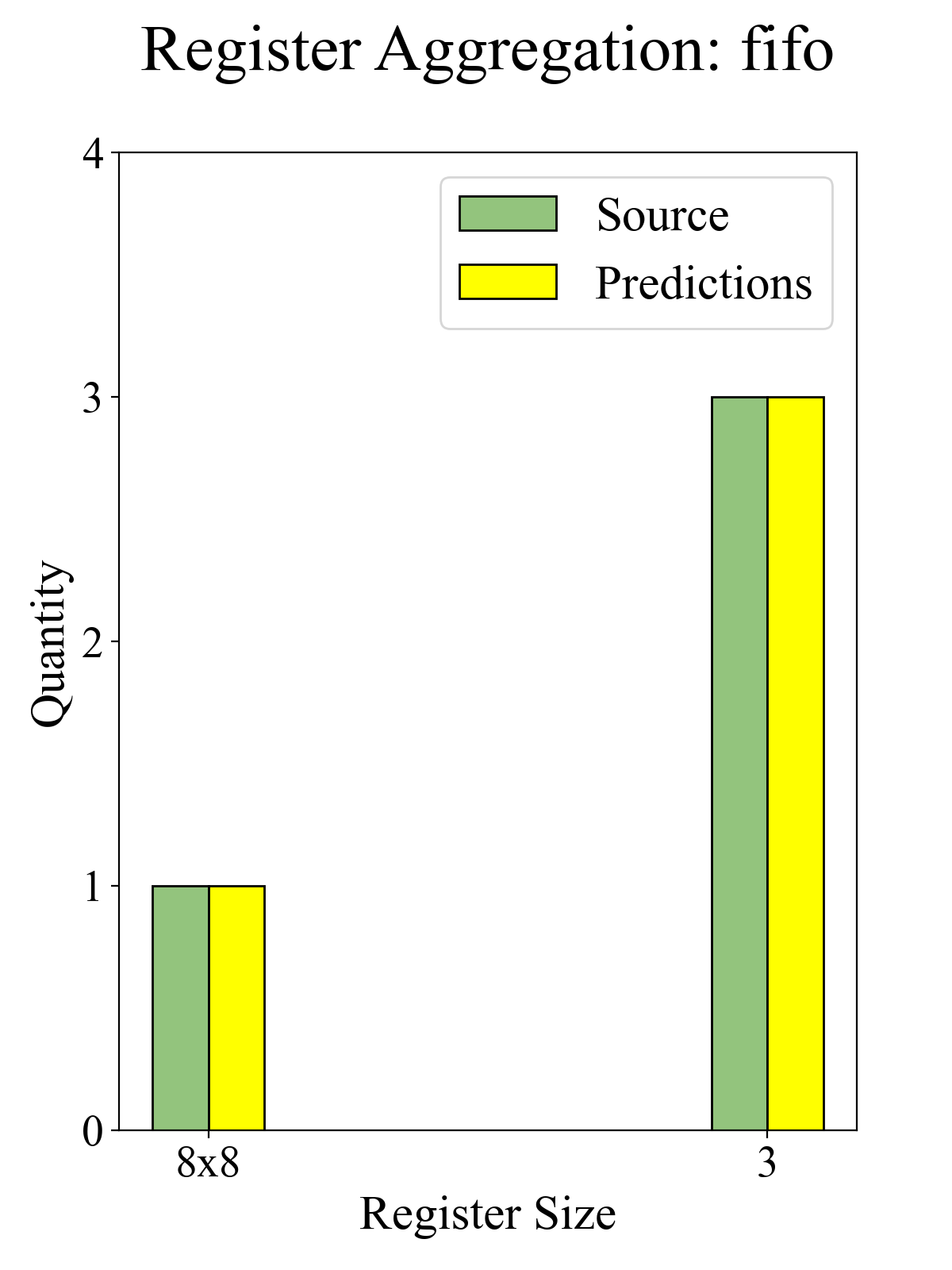}};
\node at (13.7, 0) {\includegraphics[scale=0.28]{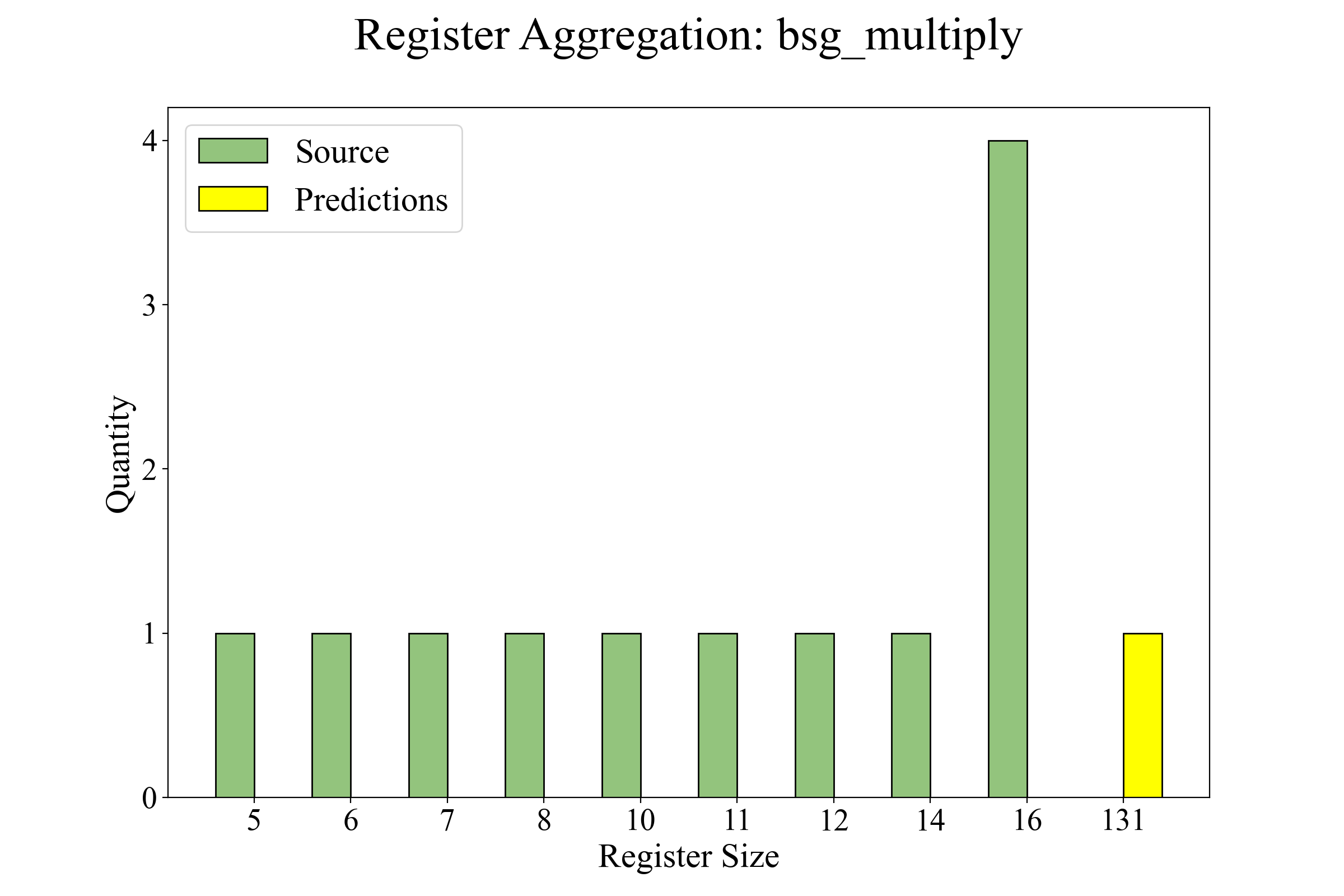}};
\node[thick,circle,draw=none,align=center] at (-2.3,-2.45) {a.};
\node[thick,circle,draw=none,align=center] at (2.7,-2.45) {b.};
\node[thick,circle,draw=none,align=center] at (9.6,-2.45) {c.};
\end{tikzpicture}
\caption{Three levels of additional aggregation. a) Source code directly matches the predicions. b) Some source code registers additionally aggregated: \: \: \: two 1-bit $\to$ one 2-bit register, two 32-bit + one 1-bit $\to$ one 65-bit register. c) All source code registers additionally aggregated into 1 131-bit register.}
\end{figure*}
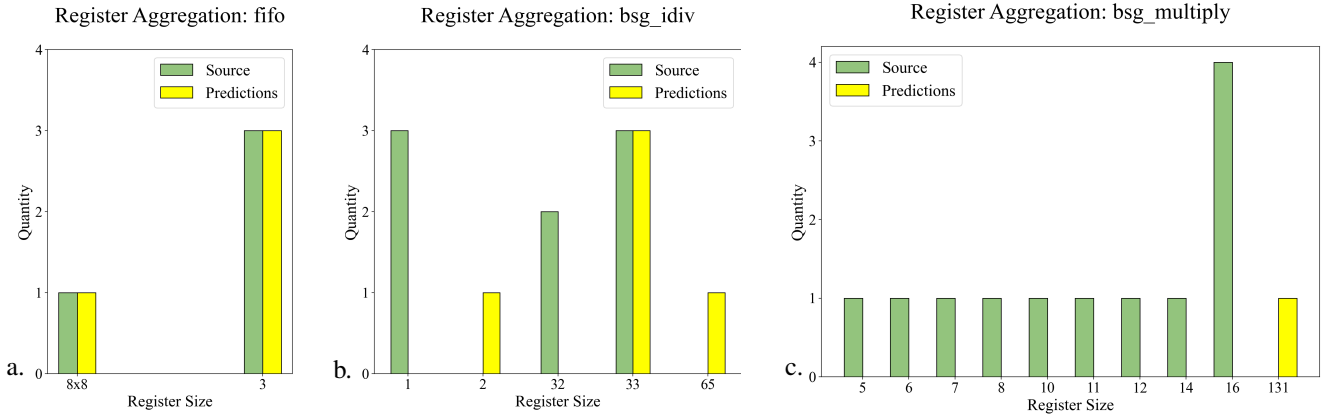

For the piso, bsg\_assembler, and bsg\_idiv (Figure 5b) examples, the technique not only recovers all the source code registers as desired, but it additionally aggregates subsets of them into larger registers. Each such subset shares an enable and can therefore be concatenated into a larger register without affecting the circuit’s behavior. We verify that the technique is indeed accurate by confirming the presence of common enable pins between multi-bit registers instantiated in the source code. This additional register aggregation occurs more dramatically in the bsg\_multiply (Figure 5c) and bsg\_multiply\_array examples, where all the source code registers get combined into one register because they all share one common enable. The additional register aggregation performed by our technique shows that our work can improve the register groupings in a netlist beyond what is originally in the source code, suggesting organizational and compaction capabilities. These capabilities open up potential new applications for decompilation, such optimizing register and memory block usage in existing HDL code.

Additional register aggregation seems to occur with the opdb\_pico example as well, but the scale is too large to identify where each source code register ends up in the register predictions. Moreover, opdb\_pico also has two memory blocks, and the technique successfully recovers the $32\times 32$  memory block but fails on the $4 \times 32$ one, outputting it as 4 32-bit registers instead. Here, the neglected memory block is still compacted into 4 registers from 128 DFFs present after synthesis, so some organizational information is retained.

For the nerv example, the technique groups all the registers correctly, additionally aggregating two 5-bit registers into a 10-bit one. However, it fails to recover the $32 \times 32$ memory block, instead outputting it as a 32 32-bit registers. The reason that this memory block is not recovered is because it has multiple read ports and addresses, breaking our memory block port assumptions and preventing the technique from finding a single address. In investigating this failure, we find that the technique does group the 32 registers with an enable before quitting the aggregation process after a unique address is not found. 

Finally, for the bsg\_cache example, additional aggregation is extended even further, with 8 memory blocks being created out of source code without any memory block instantiations. This is of great significance because 16384 DFFs after synthesis are not only recovered into 2048 8-bit registers in the source code, but compacted even further into 8 $8 \times 256$ memory blocks. This compaction from 2048 source code artifacts down to 8 artifacts does not affect the circuit's behavior while improving its organization tremendously. 

Given limitations of our current technique, such as reliance on PyRTL and memory block port assumptions, future directions include building a more language-agnostic technique and supporting multiple read or write ports in memory blocks. Moreover, additional aggregation, particularly its extension to forming new memory blocks as observed in bsg\_cache, illuminates the possibility of utilizing hardware decompilation to build implicit memory blocks from existing HDL code. To further affirm the real-world relevance of register aggregation, previously stated applications such as technology retargeting could be tested in later work.

\subsection{Runtime Analysis}
\begin{figure} 
\centering
\begin{tikzpicture}[scale=0.87]
\node at (0,0) {\includegraphics[scale=0.29]{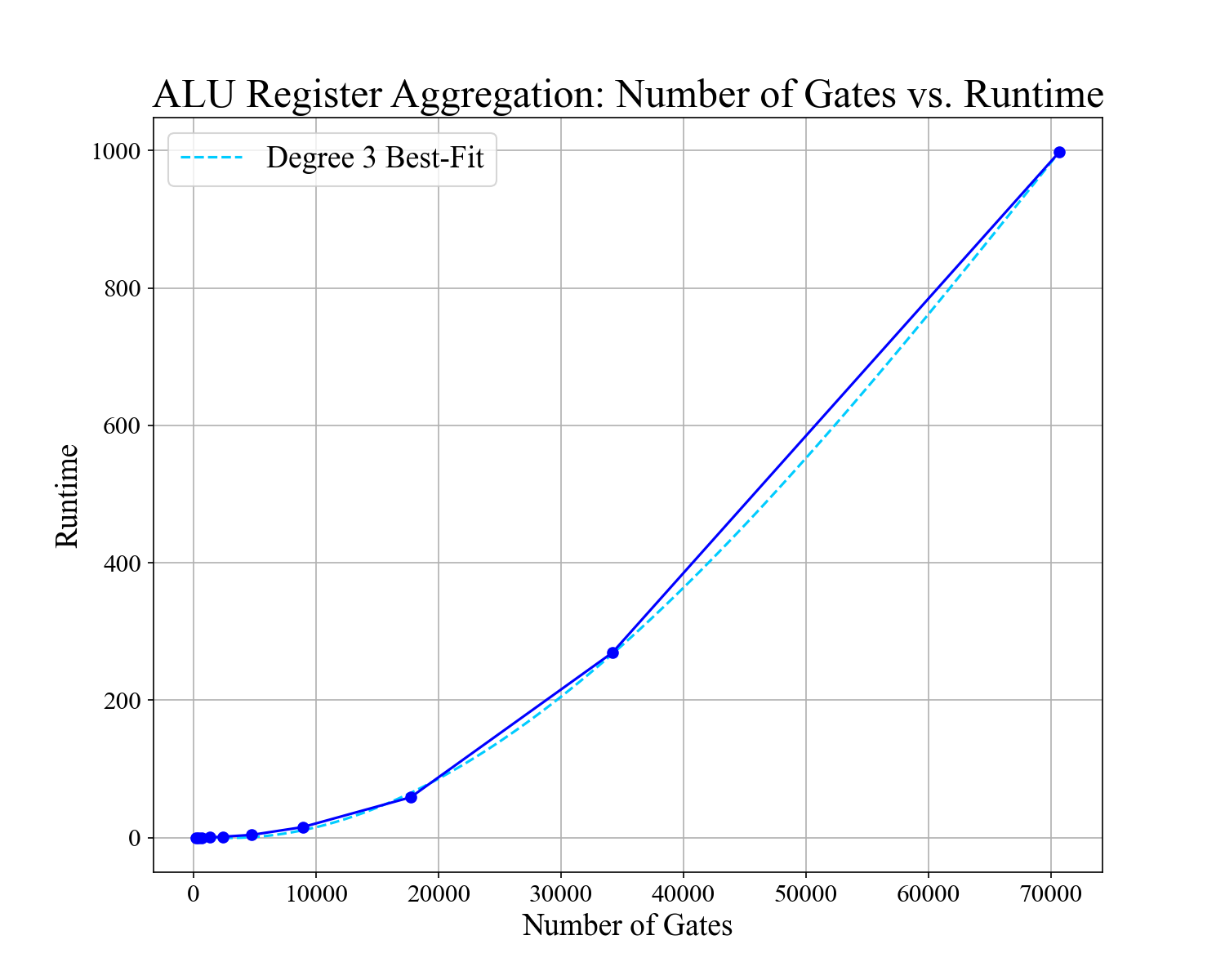}};
\node at (0,-6.7) {\includegraphics[scale=0.21]{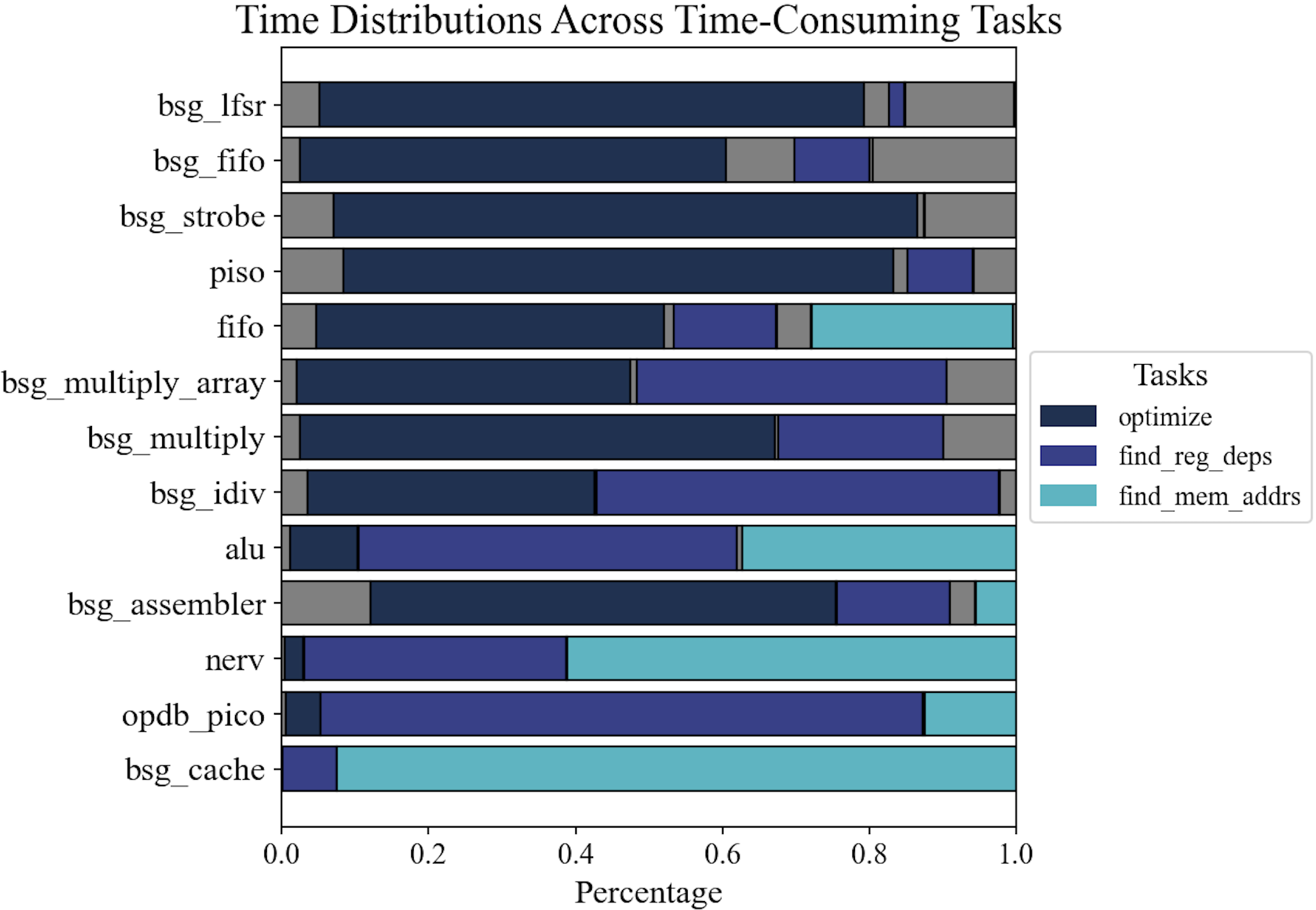}};
\node[thick,circle,draw=none,align=center] at (-5,-2.5) {a.};
\node[thick,circle,draw=none,align=center] at (-5,-2.5 - 6.7) {b.};
\end{tikzpicture}
\caption{In a), the ALU aggregation runtime with respect to netlist size is compared to a best-fit cubic polynomial. In b), the percentage of the runtime which each of three time-consuming tasks (external netlist optimization, finding register dependencies, and finding memory block addresses) takes is graphed for all of the benchmarks.}
\end{figure}

To analyze the time complexity of our aggregation technique, we measure its runtime on the alu example, varying the dimensions of the alu memory block to change the size of the netlist. As the number of gates increases, the runtime scales at a cubic rate, as seen in Figure 5a. The cubic complexity is due to the graph traversals necessary to find dependencies for each located DFF. It turns out that the number of DFFs in a netlist is roughly linear with respect to the netlist's size. As a result, register aggregation is necessarily polynomial time, meaning our current technique is reasonably fast.

We also investigate the sub processes inside the aggregation technique to see which tasks consume the most time. Given that most of the steps are linear time operations, we only measure three main graph traversal tasks. The first task is optimize, which is a method outside of the to the actual aggregation that eliminates redundancies in the netlist. The other two tasks are finding register dependencies and memory addresses, as described in Sections 2a and 2b.
    
In benchmarks with fewer netlist gates, optimize is the slowest task, but the other two tasks take up all most all of the time for larger benchmarks. This elucidates that optimize performs its necessary graph traversal operations in less than cubic time relative to the netlist size, revealing the potential to speed up other graph traversal operations in the technique to be on par with optimize. Moreover, when there are memory blocks found, memory address computation quickly outweighs finding register dependencies. This is because memory blocks are two-dimensional while registers are one-dimensional, so they have an extra factor of time complexity.

\section{\raggedright \textbf{CONCLUSION}}
Our research presents and evaluates a technique to perform register aggregation for hardware decompilation, a facet of the problem which has not been targeted before. We find that our technique is largely successful at the task of register aggregation, recovering memory elements in all 13 examples and producing accurate dimensions in 10 of them. It is also relatively efficient given its predicted cubic time complexity. Our register aggregation work expands the scope of hardware decompilation, paving the way towards applications like technology retargeting and decompiling large processors.

\section*{\raggedright \textbf{ACKNOWLEDGEMENTS}}

We would like to thank Prof. Jonathan Balkind for guidance and discussions about the research, as well as Pranjali Jain for many tips and helpful feedback. We thank Dr. Lina Kim for instruction and making this research possible through the UCSB Research Mentorship Program.

\section*{\raggedright \textbf{REFERENCES}}

\begin{enumerate}
\renewcommand{\labelenumi}{[\theenumi]}
\item \label{b1} Y. Herklotz and J. Wickerson, “Finding and understanding bugs in FPGA synthesis tools,” \emph{Proceedings of the 2020 ACM/SIGDA International Symposium on Field-Programmable Gate Arrays}, Feb. 2020. doi:10.1145/3373087.3375310
\item \label{b2} R. Nigam et al., “Predictable accelerator design with time-sensitive affine types,” \emph{Proceedings of the 41st ACM SIGPLAN Conference on Programming Language Design and Implementation}, Jun. 2020. doi:10.1145/3385412.3385974
\item \label{b3} G. H. Smith et al., “FPGA technology mapping using sketch-guided program synthesis,” \emph{Proceedings of the 29th ACM International Conference on Architectural Support for Programming Languages and Operating Systems}, Volume 2, Apr. 2024. doi:10.1145/3620665.3640387
\item \label{b4} S. Beamer, “A case for accelerating software RTL simulation,” \emph{IEEE Micro}, vol. 40, no. 4, pp. 112–119, Jul. 2020. doi:10.1109/mm.2020.2997639
\item \label{b5} M. Ganai and A. Kuehlmann, “On-the-Fly Compression of Logical Circuits,” \emph{International Workshop on Logic Synthesis}, Jul. 2000.
\item \label{b6} Z. D. Sisco, J. Balkind, T. Sherwood, and B. Hardekopf, “Loop rerolling for hardware decompilation,” \emph{Proceedings of the ACM on Programming Languages}, vol. 7, no. PLDI, pp. 420–442, Jun. 2023. doi:10.1145/3591237
\item \label{b7} G. H. Smith et al., “There and Back Again: A Netlist's Tale with Much Egraphin',” arXiv:2404.00786 [cs.AR], Mar. 2024. \\ doi:10.46586/tches.v2020.i4.309-336 
\item \label{b8} N. Albartus, M. Hoffmann, S. Temme, L. Azriel, and C. Paar, “Dana Universal Dataflow Analysis for gate-level netlist reverse engineering,” \emph{IACR Transactions on Cryptographic Hardware and Embedded Systems}, pp. 309–336, Aug. 2020. doi:10.46586/tches.v2020.i4.309-336
\item \label{b9} J. Portillo, T. Meade, J. Hacker, S. Zhang, and Y. Jin, “RERTL: Finite State Transducer Logic Recovery at Register Transfer Level,” \emph{2019 Asian Hardware Oriented Security and Trust Symposium (AsianHOST)}, Dec. 2019. doi:10.1109/asianhost47458.2019.9006699
\item \label{b10} T. Zhang, J. Wang, S. Guo, and Z. Chen, “A comprehensive FPGA reverse engineering tool-chain: From bitstream to RTL code,” \emph{IEEE Access}, vol. 7, pp. 38379–38389, 2019. doi:10.1109/access.2019.2901949
\item \label{b11} A. Gupta and A. L. Fisher, "Representation and symbolic manipulation of linearly inductive Boolean functions," \emph{Proceedings of 1993 International Conference on Computer Aided Design (ICCAD)}, Santa Clara, CA, USA, 1993, pp. 192-199.\\ doi: 10.1109/ICCAD.1993.580055
\item \label{b12} UC Berkeley. Berkeley logic interchange format (BLIF). \emph{Oct Tools Distribution}, vol. 2, pp. 197–247, Jul. 1992.
\item \label{b13} J. Clow et al., “A pythonic approach for Rapid Hardware Prototyping and instrumentation,” \emph{2017 27th International Conference on Field Programmable Logic and Applications (FPL)}, Sep. 2017. doi:10.23919/fpl.2017.8056860
\item \label{b14} M. B. Taylor, “Basejump STL: Systemverilog needs a standard template library for hardware design,” \emph{Proceedings of the 55th Annual Design Automation Conference}, Jun. 2018. doi:10.1145/3195970.3199848
\item \label{b15} G. Tziantzioulis et al., “OPDB: A Scalable and Modular Design Benchmark,” \emph{IEEE Transactions on Computer-Aided Design of Integrated Circuits and Systems}, vol. 41, no. 6, pp. 1878–1887, Jun. 2022. doi:10.1109/tcad.2021.3096794
\end{enumerate}

\end{document}